\documentclass[prd,twocolumn,floatfix]{revtex4-1}
\usepackage{dcolumn,amsmath}
\usepackage{graphicx}
\usepackage{bm}
\usepackage{amssymb}
\usepackage{hyperref}
\usepackage{multirow}
\usepackage{footnote}
\usepackage{slashed}
\usepackage{xcolor}
\usepackage{pdfcomment}
\usepackage{commath}
\usepackage{subcaption}
\usepackage{tikz}
\usepackage[font=small,labelfont=bf,
   justification=justified,
   format=plain]{caption}
\usetikzlibrary{positioning,arrows}
\usetikzlibrary{decorations.pathmorphing}
\usetikzlibrary{decorations.markings}
\usepackage{tikz-feynman,contour}
\usepackage{diagbox}
\usepackage{cancel}
\usepackage{mathrsfs}
\usepackage{enumitem}

\newcommand{\RomanNumeralCaps}[1]
{\MakeUppercase{\romannumeral #1}}

\newcommand{\RUG}{
Van Swinderen Institute for Particle Physics and Gravity,
\\University of Groningen, Nijenborgh 4, 9747AG Groningen, The Netherlands\\
}

\begin{document}

\title{Connection between $\nu n \rightarrow \bar{\nu} \bar{n}$ reactions and $n$-$\bar{n}$ oscillations via additional Higgs triplets}

\author{Yongliang Hao}
\email{y.hao@rug.nl}

\affiliation{\RUG}
\date{\today}

\begin{abstract}

In this work, we investigate the connection and compatibility between $\nu n \rightarrow \bar{\nu} \bar{n}$ reactions and $n$-$\bar{n}$ oscillations based on the $SU(3)_c \times SU(2)_L \times U(1)$ symmetry model with additional Higgs triplets. We explore the possibility that the scattering process $\nu n\rightarrow \bar{\nu}\bar{n}$ produced by low-energy solar neutrinos gives rise to an unavoidable background in the measurements of $n$-$\bar{n}$ oscillations. We focus on two different scenarios, depending on whether the $(B-L)$ symmetry could be broken. We analyze the interplay of the various constraints on the two processes and their observable consequences. In the scenario where both $(B+L)$ and $(B-L)$ could be broken, we point out that if all the requirements, mainly arising from the type-\RomanNumeralCaps{2} seesaw mechanism, are satisfied, the parameter space would be severely constrained. In this case, although the masses of the Higgs triplet bosons could be within the reach of a direct detection at the LHC or future high-energy experiments, the predicted $n$-$\bar{n}$ oscillation times would be completely beyond the detectable regions of the present experiments. In both scenarios, the present experiments are unable to distinguish a $\nu n \rightarrow \bar{\nu} \bar{n}$ reaction event from a $n$-$\bar{n}$ oscillation event within the accessible energy range. Nevertheless, if any of the two processes is detected, there could be signal associated with new physics beyond the Standard Model.


\end{abstract}

\maketitle

\section{Introduction} \label{section1}

Baryon number ($B$) and lepton number ($L$) are usually considered as accidental symmetries in three fundamental interactions of the Standard Model (SM) \cite{Cerdeno2019ngx}. Some non-perturbative effects in the SM may violate the $B$, $L$, and $(B+L)$ symmetries, but the difference $(B-L)$ is still conserved \cite{hooft1976symmetry,hooft1976computation,arnold2013simplified,ellis2016search}. $B$-violation, in particular, is one of the three criteria suggested by Sakharov to explain the observed matter-antimatter asymmetry in our Universe \cite{sakharov1967violation}. Additionally, in order to generate the observed asymmetry, the $(B-L)$ symmetry must be conserved too, or else the non-perturbative sphaleron process may smooth out such asymmetry \cite{morrissey2012electroweak,fujita2016large}. In some new physics models such as the left-right $SU(3)_c \times SU(2)_L \times SU(2)_R \times U(1)_{B-L}$ symmetry model \cite{pati1974lepton,pati1975erratum,Mohapatra1974hk,Senjanovic1975rk}, the grand unified $SU(5)$ symmetry model \cite{georgi1974unity}, the partially unified $SU(4)_c \times SU(2)_L \times SU(2)_R$ symmetry model \cite{mohapatra1980local,babu2009neutrino,davidson1979b} etc., unlike $B$ alone or $L$ alone, the difference $(B-L)$ is implemented as a symmetry in describing the interactions among quarks and leptons, predicting the existence of the $(B+L)$-violating processes such as hydrogen-antihydrogen ($H$-$\bar{H}$) oscillations \cite{mohapatra1982hydrogen,mohapatra1983spontaneous,mohapatra1983higgs} and $\nu n \rightarrow \bar{\nu} \bar{n}$ reactions. In such models, symmetry can be broken spontaneously to the $SU(3)_c \times SU(2)_L \times U(1)$ symmetry model, leading to $n$-$\bar{n}$ oscillations \cite{mohapatra1980local,mohapatra1982hydrogen,mohapatra1983spontaneous,mohapatra1983higgs,babu2009neutrino} and neutrino Majorana masses \cite{mohapatra1983spontaneous}.
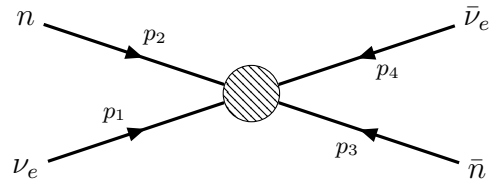
\begin{figure}[b]
  \centering
  \begin{tikzpicture}
    \begin{feynman}
      \vertex[blob,label={right:$ $}] (m) at ( 0, 0) {\contour{white}{}};
      \vertex (a) at (-3,-1.0) {\large $\nu_e$};
      \vertex (b) at ( 3,-1.0) {\large $\bar{n}$};
      \vertex (c) at (-3, 1.0) {\large $n$};
      \vertex (d) at ( 3, 1.0) {\large $\bar{\nu}_e$};
      \diagram* {
        (a) -- [fermion,very thick,edge label=$p_1$] (m), (c) -- [fermion,very thick,edge label=$p_2$] (m),
        (b) -- [fermion,very thick,edge label=$p_3$] (m), (d)-- [fermion,very thick,edge label=$p_4$] (m),
      };
    \end{feynman}
  \end{tikzpicture}
\caption{A neutron is scattered by a neutrino changing into an antineutron and an antineutrino (the $\nu n\rightarrow \bar{\nu}\bar{n}$ reaction process). Here $p_1$ and $p_2$ are the four-momenta of the incoming neutrino and neutron respectively, and $p_4$ and $p_3$ are the four-momenta of the outgoing antineutrino and antineutron respectively.}
\label{Figfeyn}
\end{figure}
In cosmology, some leptogenesis scenarios are proposed to explain the asymmetry between matter and antimatter, but if the $B$-violating $n$-$\bar{n}$ oscillations are observed, then the leptogenesis models will be ruled out, assuming that it occurs at the energy scale where $n$-$\bar{n}$ oscillations are in equilibrium \cite{mohapatra2009neutron,dutta2006observable}. Furthermore, previous studies \cite{mohapatra1982hydrogen,mohapatra1983spontaneous,mohapatra1983higgs} show that it is possible to estimate the $H$-$\bar{H}$ oscillation time by comparing it with the $n$-$\bar{n}$ oscillation time, where a large degree of uncertainty could be eliminated \cite{mohapatra1982hydrogen,mohapatra1983spontaneous,mohapatra1983higgs} and the prediction power can be greatly improved. On the other hand, it is considered that the $(B+L)$ symmetry is anomalous \cite{oosterhof2019baryon}. In some other extensions to the SM, the breaking of $(B+L)$ symmetry is also introduced as an important feature \cite{cerdeno2018bl,Cerdeno2019ngx}. Therefore, testing such global symmetries could signal new physics beyond the SM \cite{phillips2016neutron,cerdeno2018bl}.

The change of a neutron into an antineutron, namely neutron-antineutron ($n$-$\bar{n}$) oscillations, violates $B$, $(B+L)$, and $(B-L)$ by two units ($|\Delta B|=2$, $|\Delta (B+L)|=2$, and $|\Delta (B-L)|=2$). The results of the searches for $n$-$\bar{n}$ oscillations have been presented by numerous experiments in different mediums \cite{phillips2016neutron} such as field-free vacuum, bound states, as well as external fields. On the one hand, however, no significant evidence has been observed for $n$-$\bar{n}$ oscillations so far. The lower limits on the $n$-$\bar{n}$ oscillation times for neutrons inside nuclei are reported by various experiments, such as Irvine-Michigan-Brookhaven (IMB) \cite{jones1984search}, Kamiokande (KM) \cite{takita1986search}, Frejus \cite{berger1990search}, Soudan-2 (SD-2) \cite{chung2002search}, Super-Kamiokande (Super-K) \cite{abe2015search}, and Sudbury Neutrino Observatory (SNO) \cite{aharmim2017search}. In field-free vacuum, the present best lower limit on the $n$-$\bar{n}$ oscillation time is reported by the ILL experiment \cite{baldo1994new}.

At the quark level, $B$, $(B+L)$, and $(B-L)$ violations can only be described by high dimensional operators associated with some large mass scales and thus the effect is greatly suppressed and usually considered to be undetectable at low energies \cite{nieves1984analysis}. The lower limits on $n$-$\bar{n}$ oscillation times for neutrons in matter are derived from the stability of nuclei \cite{jones1984search,takita1986search,chung2002search,berger1990search,abe2015search,aharmim2017search}. However, the instability of nuclei induced by external low-energy solar neutrinos has not been excluded because of the detector thresholds. In the presence of low-energy solar neutrinos, we will see that the present detectors are unable to distinguish a $\nu n \rightarrow \bar{\nu} \bar{n}$ reaction event from a $n$-$\bar{n}$ oscillation event. Therefore, it is reasonable to assume that some of the reported $n$-$\bar{n}$ oscillation candidates may actually be produced by low-energy solar neutrinos in the scattering process $\nu n\rightarrow \bar{\nu}\bar{n}$ as depicted in Fig. \ref{Figfeyn}. Following previous studies of the $H$-$\bar{H}$ oscillations in Refs. \cite{mohapatra1982hydrogen,mohapatra1983spontaneous,mohapatra1983higgs}, it is also possible to relate $\nu n \rightarrow \bar{\nu} \bar{n}$ reactions to $n$-$\bar{n}$ oscillations, meanwhile eliminating a large degree of uncertainty. In this work, we explore the possible connection between $n$-$\bar{n}$ oscillations and $\nu n \rightarrow \bar{\nu} \bar{n}$ reactions based on the $SU(3)_c \times SU(2)_L \times U(1)$ symmetry model with additional Higgs triplets. As it will be shown in the following sections, although, currently, there is no information available on the experimental rate for the $\nu n \rightarrow \bar{\nu} \bar{n}$ reaction process, the ratio of the interaction rate for $\nu n \rightarrow \bar{\nu} \bar{n}$ reaction to the interaction rate for $n$-$\bar{n}$ oscillation can be estimated from a theoretical point of view by connecting the two processes using the Higgs triplet and neutrino masses. In such an approach, some parameters that appear both in the numerator and in the denominator, such as the nuclear suppression factor, can be eliminated, making it possible to place constraints on the two processes and analyze their observable consequences using the results of the searches for $n$-$\bar{n}$ oscillations. Throughout the paper, if not otherwise mentioned, we only consider the first generation of particles and antiparticles, and thus all neutrinos under discussion are electron-type neutrinos ($\nu \equiv \nu_e$).

\section{The Model} \label{section2}

\begin{figure}[b]
\begin{minipage}[b]{4cm}
\centering
\begin{tikzpicture}
    \begin{feynman}
      \vertex[dot] (o) at (0.0,0.0) {\large $o$};
      \vertex (a) at (1.0,0.0) {\large $\langle \Delta_{\nu \nu} \rangle$};
      \vertex (a1) at (2.0,-1.0) {\large $ $};
      \vertex (a2) at (2.0,1.0) {\large $ $};
      
      \vertex (b) at (0.0,1.0);
      \vertex (b1) at (1.0,2.0) {\large $d$};
      \vertex (b2) at (-1.0,2.0) {\large $\bar{d}$};
      
      \vertex (c) at (-1.0,0.0);
      \vertex (c1) at (-2.0,1.0) {\large $u$};
      \vertex (c2) at (-2.0,-1.0) {\large $\bar{u}$};
      
      \vertex (d) at (0.0,-1.0);
      \vertex (d1) at (-1.0,-2.0) {\large $d$};
      \vertex (d2) at (1.0,-2.0) {\large $\bar{d}$};
      
      \diagram* {
      (b1) -- [fermion,very thick,edge label=$ $] (b), 
      (b2) -- [fermion,very thick,edge label=$ $] (b),
      (c1) -- [fermion,very thick,edge label=$ $] (c), 
      (c2) -- [fermion,very thick,edge label=$ $] (c), 
      (d1) -- [fermion,very thick,edge label=$ $] (d), 
      (d2) -- [fermion,very thick,edge label=$ $] (d),
      (a) -- [scalar,very thick,edge label={\large $ $}] (o), 
      (c) -- [scalar,very thick,edge label={\large $\Delta_{uu}$}] (o),
      (b) -- [scalar,very thick,edge label={\large $\Delta_{dd}$}] (o),
      (d) -- [scalar,very thick,edge label={\large $\Delta_{dd}$}] (o)
      };
    \end{feynman}
    \label{figa}
\end{tikzpicture}
\subcaption{$n$-$\bar{n}$ oscillation}
\end{minipage}
\begin{minipage}[b]{4cm}                 
\centering
\begin{tikzpicture} 
    \begin{feynman}
      \vertex[dot] (o) at (0.0,0.0) {\large $o$};
      \vertex (a) at (1.0,0.0);
      \vertex (a1) at (2.0,-1.0) {\large $\nu_e$};
      \vertex (a2) at (2.0,1.0) {\large $\bar{\nu_e}$};
      
      \vertex (b) at (0.0,1.0);
      \vertex (b1) at (1.0,2.0) {\large $d$};
      \vertex (b2) at (-1.0,2.0) {\large $\bar{d}$};
      
      \vertex (c) at (-1.0,0.0);
      \vertex (c1) at (-2.0,1.0) {\large $u$};
      \vertex (c2) at (-2.0,-1.0) {\large $\bar{u}$};
      
      \vertex (d) at (0.0,-1.0);
      \vertex (d1) at (-1.0,-2.0) {\large $d$};
      \vertex (d2) at (1.0,-2.0) {\large $\bar{d}$};
      
      \diagram* {
      (a1) -- [fermion,very thick,edge label=$ $] (a), 
      (a2) -- [fermion,very thick,edge label=$ $] (a),
      (b1) -- [fermion,very thick,edge label=$ $] (b), 
      (b2) -- [fermion,very thick,edge label=$ $] (b),
      (c1) -- [fermion,very thick,edge label=$ $] (c), 
      (c2) -- [fermion,very thick,edge label=$ $] (c), 
      (d1) -- [fermion,very thick,edge label=$ $] (d), 
      (d2) -- [fermion,very thick,edge label=$ $] (d),
      (a) -- [scalar,very thick,edge label={\large $\Delta_{\nu \nu}$}] (o),
      (c) -- [scalar,very thick,edge label={\large $\Delta_{uu}$}] (o),
      (b) -- [scalar,very thick,edge label={\large $\Delta_{dd}$}] (o),
      (d) -- [scalar,very thick,edge label={\large $\Delta_{dd}$}] (o)
      };
    \end{feynman}
    \label{figb}
\end{tikzpicture}
\subcaption{$\nu n\rightarrow\bar{\nu} \bar{n}$ reaction}
\end{minipage}
\caption{Possible diagrams for (a) $n$-$\bar{n}$ oscillations and (b) $\nu n\rightarrow\bar{\nu} \bar{n}$ reactions mediated by additional Higgs triplets, namely diquarks and dileptons.}
\label{figab}
\end{figure}
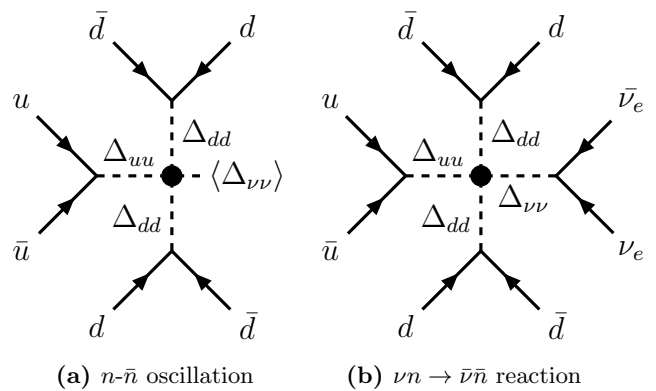

Fig. \ref{figab} (a) and Fig. \ref{figab} (b) show the possible diagrams at the quark level for $n$-$\bar{n}$ oscillations and $\nu n\rightarrow\bar{\nu} \bar{n}$ reactions respectively mediated by Higgs triplet particles \cite{mohapatra1982hydrogen,mohapatra1983spontaneous}. The two processes can be described by the interactions based on the $SU(3)_c \times SU(2)_L \times U(1)$ symmetry model with enlarged Higgs sector, which can be embedded in some grand (or partially) unified models with higher symmetries. In this model, the fermionic fields take the following conventional form \cite{mohapatra1982hydrogen,mohapatra1983spontaneous},
\begin{equation}
\begin{split}
Q_L\Bigl(3, 2,\frac{1}{3}\Bigr) &
=
\left(
  \begin{array}{c}
   u\\
   d
  \end{array}
  \right)_L, \quad
\Psi_L\Bigl(1, 2,-1\Bigr)=
\left(
  \begin{array}{c}
   \nu\\
   e
  \end{array}
  \right)_L \\
u_R\Bigl(3, 1, \frac{4}{3}\Bigr) &, \quad d_R\Bigl(3, 1, -\frac{2}{3}\Bigr), \quad e_R\Bigl(1, 1, -2\Bigr).
\end{split}
\end{equation}
Here, the right and left handed spinors are defined as $\psi_{R/L}\equiv P_{R/L} \psi$, where $P_{R/L}\equiv(1\pm \gamma^5)/2$ are the right and left chiral projection operators. In addition to the $SU(2)_L$ Higgs doublet, two additional $SU(2)_L$ Higgs triplets are incorporated into the model as follows \cite{mohapatra1982hydrogen,mohapatra1983spontaneous}
\begin{equation}
\Phi \Bigl(1,2,1\Bigr), \quad \Delta_q \Bigl(\bar{6},3,-\frac{2}{3}\Bigr), \quad \Delta_l \Bigl(1,3,2\Bigr).
\end{equation}
Here, $\Phi \equiv (\phi^+, \phi_0)^T$ is the Higgs doublet, while $\Delta_q$ and $\Delta_l$ are the two newly added Higgs triplets, namely diquarks and dileptons, which can be written in the following matrix form \cite{nieves1984analysis,chen2011type,de2019implementing},
\begin{equation}
\begin{split}
\Delta_q &
=
\left(
  \begin{array}{cc}
   \frac{\Delta_{ud}}{\sqrt{2}}         &  \Delta_{dd}\\
   \Delta_{uu} & -\frac{\Delta_{ud}}{\sqrt{2}}
  \end{array}
  \right)
\end{split}
\end{equation}
\begin{equation}
\begin{split}
\Delta_l &
=
\left(
  \begin{array}{cc}
   \frac{\Delta_{\nu e}}{\sqrt{2}}         &  \Delta_{ee}\\
   \Delta_{\nu \nu} & -\frac{\Delta_{\nu e}}{\sqrt{2}}
  \end{array}
  \right)
\end{split}
\end{equation}
As argued in Ref. \cite{mohapatra1983spontaneous}, in this model, the corresponding Higgs potential can be chosen to preserve a discrete symmetry so that the compatibility with the current experimental constraints on the proton lifetime $\tau_p \gtrsim 10^{31}$-$10^{33}$ yr \cite{tanabashi2018review}, which is model dependent, is assured.

The set of relevant operators, responsible for the two processes depicted in Fig. \ref{figab}, can be chosen as \cite{mohapatra1980local,barbieri1981spontaneous,mohapatra1982hydrogen,mohapatra1983spontaneous,babu2012coupling},
\begin{equation}
\begin{split}
O_{s}  \equiv & \quad g_{\alpha \beta} Q^T_{\alpha L} C^{-1} i \sigma_2 \Delta_q Q_{\beta L} + f_{\alpha \beta} \Psi^T_{\alpha L} C^{-1} i \sigma_2 \Delta_l \Psi_{\beta L}\\
                & + \lambda \epsilon_{ikm} \epsilon_{jln} \Delta_{dd}^{ij} \Delta_{dd}^{kl} \Delta_{uu}^{mn} \Delta_{\nu \nu} + \text{H.c.}
\end{split}
\label{fgdefinition}
\end{equation}
Here, $i,j,k,l,m,n$ stand for $SU(3)_c$ indices, and $\alpha$,$\beta$ stand for $SU(2)_L$ indices. The parameters $g_{\alpha \beta}$, $f_{\alpha \beta}$ and $\lambda$ are the vertex coupling constants in the Yukawa and gauge sectors. $C$ represents charge conjugation operator. Actually, the two diagrams in Fig. \ref{figab} are not the only diagrams describing the interactions responsible for the two processes, whereas the interactions could also be possibly mediated by the $\Delta_{ud}$ boson \cite{vergados1982study,babu2012coupling,babu2009neutrino,patraa2014post}. For simplicity, in the following discussions, without loss of generality, we only focus on the interactions depicted in Fig. \ref{figab} but the conclusions could be applied to the interactions mediated by the $\Delta_{ud}$ boson, if we assume that all the Higgs triplet components have the same mass.

Since the solar neutrinos have very low energies, it is reasonable to describe the $\nu n\rightarrow\bar{\nu} \bar{n}$ reaction process using an effective Lagrangian at the hadron level. The four-fermion contact interaction was first proposed as an effective field theory in describing $\beta$-decay \cite{konopinski1955fermi} at low energies, where neutrons and protons are treated as point particles. The solar neutrinos have an average energy of around $E_{\nu} \simeq 0.53$ MeV \cite{ianni2014solar}, and the corresponding wavelength is so long that in general they cannot probe the structure of the nucleons. The degrees of freedom can be chosen as neutrons and neutrinos. Therefore, in the energy range of solar neutrinos, the contact interaction is supposed to be applicable \cite{feinberg1978multiplicative,bramante2015proton,grossman2018revisiting} without considering the structure of the neutron.

We assume that the effective Lagrangian at the hadron level, which describes the $\nu n \rightarrow \bar{\nu} \bar{n}$ reaction process (depicted in Fig. \ref{Figfeyn}) via scalar contact interactions, takes the following form,
\begin{equation}
-\mathscr{L}_{b}^{eff} \equiv G_b |\psi_q (0)|^4 \Bigl( \bar{n}^c_R \nu_L \Bigr) \Bigl( \bar{\nu}^c_R n_L \Bigr)\\
\label{lagrangian}
\end{equation}
Here, $\psi_q (0)$ is the quark wave function at the origin and $|\psi_q (0)|^2 \simeq 0.0144 (3)(21)$ GeV$^3$ \cite{aoki2017improved} is given by the lattice QCD calculations, with the numbers in parenthesis being statistical and systematic uncertainties. $G_b$ is the effective coupling constant and will be discussed in more detail in the following sections. The superscript $c$ represents charge conjugation, and the scalar interaction couples states with opposite chirality.

The constraints on nucleon instability can be determined through the measurements of two decay modes, such as $n$-$\bar{n}$ oscillations \cite{jones1984search,takita1986search,berger1990search,chung2002search,abe2015search,aharmim2017search} and the dineutron decay: $n n \rightarrow \bar{\nu} \bar{\nu}$ \cite{bernabei2000search}. Such decay modes violate $B$ and $(B+L)$ but the dineutron decay preserves $(B-L)$. Both the $n$-$\bar{n}$ oscillation process and the $\nu n\rightarrow\bar{\nu} \bar{n}$ reaction process lead to the change of a neutron into an antineutron, followed by antineutron annihilation with the surrounding nucleons into pions \cite{phillips2016neutron,oosterhof2019baryon}. However, the $n n \rightarrow \bar{\nu} \bar{\nu}$ process, which can be realized after making Fierz transformations to Eq. (\ref{lagrangian}), is featured with the decay of nucleus into two back-to-back energetic neutrinos, which are nearly invisible to detectors. The experimental limits on the lifetimes for the decay mode with electromagnetically or strongly interacting final states are several orders of magnitude larger than the ones for the decay mode with weakly interacting final states such as the dineutron decay ($n n \rightarrow \bar{\nu} \bar{\nu}$) \cite{bernabei2000search}.
For this reason, we focus on the $n$-$\bar{n}$ process (and the $\nu n\rightarrow\bar{\nu} \bar{n}$ process) rather than the $n n \rightarrow \bar{\nu} \bar{\nu}$ process in the following discussions.

Similarly, one could also construct the effective Lagrangian with $(B+L)$ violations for the charged baryon and lepton sector
\cite{bramante2015proton,grossman2018revisiting,sussman2018dinucleon}. The relevant processes are the $H$-$\bar{H}$ oscillations: $e^{-}p  \rightarrow e^{+} \bar{p}$ and the diproton decay: $pp  \rightarrow  e^{+}e^{+}$. A recent study shows that the determined constraint on the $pp \rightarrow  l^{+}l^{+}$ process has excluded new physics below an energy scale of around 1.6 TeV \cite{bramante2015proton} and the bounds on the $e^{-}p  \rightarrow e^{+}\bar{p}$ process are weaker than the ones on the $pp \rightarrow  l^{+}l^{+}$ process \cite{bramante2015proton}.

At the hadron level, the differential cross section of the $\nu n \rightarrow \bar{\nu} \bar{n}$ reaction process in the specific model given by Eq. (\ref{lagrangian}) in the center-of-mass frame can be written as \cite{nagashima2010elementary},
\begin{eqnarray}
\frac{d\sigma_b (\nu n \rightarrow \bar{\nu} \bar{n})}{d \Omega} = \frac{|\overline{M_b}|^2}{64 \pi^2 s} \frac{|p_f|}{|p_i|}
\end{eqnarray}
where $d\Omega \equiv \sin{\theta} d\theta d\phi$ with $\theta$ and $\phi$ being the scattering angles. The Mandelstam variable $s$ is defined in the usual way. In this case, it is easy to see that the relation $|p_f|=|p_i|$ holds. The effective squared amplitude $|\overline{M_b}|^2$ can be obtained by summing over all final spin configurations and averaging over all initial spin configurations: 
\begin{equation}
\begin{split}
|\overline{M_b}|^2 =& \frac{1}{2} \left( G_b |\psi_q (0)|^4 \right)^2  \text{Tr}\left[  (\cancel{p_1}+m_{\nu})  (\cancel{p_3}-m_n)  \right] \\
&\times \text{Tr} \left[  (\cancel{p_2}+m_n)  (\cancel{p_4}-m_{\nu})  \right]
\end{split}
\label{amplitude}
\end{equation}
where $m_n$ ($m_{\nu}$) is the neutron (neutrino) mass.

In this work, the cross section of the $\nu n \rightarrow \bar{\nu} \bar{n}$ reaction process is calculated using the FeynCalc package \cite{mertig1991feyn,shtabovenko2016new}. Since the solar neutrinos are mainly in the energy range from a few keV to 10 MeV \cite{wurm2017solar}, which satisfy the condition $m_\nu \ll E_\nu \ll m_n$, and thus, comparing with their energies, the tiny neutrino mass could be ignored. The corresponding cross section in the range $m_\nu \ll E_\nu \ll m_n$ can be written down as follows: 
\begin{equation}
\sigma_b(E_\nu) \simeq \frac{G_b^2 |\psi_q (0)|^8 E_{\nu}^2}{2 \pi}.
\label{eq7}
\end{equation}
where $E_{\nu}$ is the solar neutrino energy. In the following sections, we will illustrate that $\nu n \rightarrow \bar{\nu} \bar{n}$ reactions produced by high-intensity solar neutrinos could be considered as an unavoidable background in search for $n$-$\bar{n}$ oscillations.

\section{Connection between $\nu n \rightarrow \bar{\nu} \bar{n}$ reactions and $n$-$\bar{n}$ oscillations} \label{section3}

 The solar neutrinos are produced from various nuclear fusion reactions \cite{bahcall1989neutrino,giunti2007fundamentals,xing2011neutrinos}, such as the $pp$ fusion chain, the CNO cycle, etc. The corresponding fluxes can be predicted by the so-called Standard Solar Model (SSM) \cite{bahcall1962beta,bahcall1989neutrino,vinyoles2017new} and can be found from Refs. \cite{bahcall1996standard,bahcall1997gallium,abdurashitov2002solar,serenelli2007standard,serenelli2009new,serenelli2011solar,haxton2013solar,bellini2014neutrinos,bergstrom2016updated,vinyoles2017new,miramonti2018solar,agostini2017first,bellini2014final}. Numerous experiments \cite{Hampel1997fc,altmann2005complete,abdurashitov2009measurement,aharmim2013combined,bellini2014final,abe2016solar,cleveland1998measurement,hampel1999gallex,fukuda1996solar,bellini2011precision} 
have been designed to detect solar neutrinos with various thresholds (see e.g. Ref. \cite{giganti2018neutrino} or Tab. \ref{tabtwo}). The neutrino flavor oscillation has been confirmed and the corresponding ratios of the observed to expected neutrino rates for solar neutrinos in particular have also been given in various cases \cite{antonelli2013solar,agostini2017first,bellini2014final,abe2016solar}.

\begin{table*}[t]
\caption{Solar neutrino fluxes from various sources at the Earth and the corresponding signal fractions $\chi_i$ for $\nu n \rightarrow \bar{\nu} \bar{n}$ reactions. Numbers in parentheses stand for the power of 10.}
\begin{ruledtabular}
\begin{tabular}{l|ccccccccc}                                         
\diagbox{Parameter}{Neutrino}&$pp$&$^{13}N$&$^{15}O$&$^{17}F$&$^{8}B$&$hep$&$^{7}Be$(384 keV)&$^{7}Be$(862 keV)&$pep$\\\hline
Flux (cm$^{-2}$s$^{-1}$)&$5.98(10)^a$&$2.78(8)^a$&$2.05(8)^a$&$5.29(6)^a$&$5.46(6)^a$&$7.98(3)^a$&$5.30(8)^b$&$4.47(9)^b$&$1.44(8)^a$\\
Survival probability &0.542$^b$&0.528$^b$&0.517$^b$&0.517$^b$&0.384$^b$&0.30$^c$&0.537$^b$&0.524$^b$&0.514$^b$\\
Signal fraction $\chi_i$   &$53.14\%$&$1.71\%$&$2.47\%$&$0.06\%$&$2.30\%$&$0.01\%$&$0.87\%$&$36.23\%$&$3.20\%$\\
\end{tabular}
\end{ruledtabular}
\centering
{$^a$Ref. \cite{vinyoles2017new}. $^b$Ref. \cite{bellini2014final}. $^c$Ref. \cite{abe2016solar}.
}
\label{tabone}
\end{table*}

\begin{table*}[t]

\caption{Results of the searches for $n$-$\bar{n}$ oscillations inside nuclei. Such information is used to put constraints on the $\nu n \rightarrow \bar{\nu} \bar{n}$ reaction process. }
\begin{ruledtabular}
\begin{tabular}{l|ccccc}                                         
\diagbox{Parameter}{Exp.}&KM \cite{takita1986search}&Frejus \cite{berger1990search}&SD-2 \cite{chung2002search}&Super-K \cite{abe2015search} &SNO \cite{aharmim2017search}\\\hline           
Exposure (neutron$\cdot$yr)            &$3.0\times 10^{32}$   &$5.0\times 10^{32}$ &$2.19\times 10^{33}$     &$2.45\times 10^{34}$      &$2.047\times 10^{32}$     \\
Candidates $S_{0}$   &0   &0 &5        &24       &23       \\
Backgrounds $B_0$     &0.9   &2.5 &4.5      &24.1     &30.5     \\
Efficiency $\epsilon$ &0.33   &0.30 & 0.18   &0.121    &0.54    \\    
Threshold (MeV)           &7$^a$   &200$^b$ &100$^c$  &3.5$^d$      &3.5$^e$ \\
\end{tabular}
\end{ruledtabular}

\centering
{
$^a$ Ref. \cite{fukuda1996solar}. $^b$ Ref. \cite{berger1989study}. $^c$ Ref. \cite{may1986status}. $^d$ Ref. \cite{abe2016solar}. $^e$ Ref. \cite{aharmim2013combined}. 
}
\label{tabtwo}
\end{table*}

\begin{table*}[t]

\caption{Bounds on the masses of the Higgs triplet $M_{\Delta}$ arising from the results of the searches for $n$-$\bar{n}$ oscillations using acceptable vertex coupling constants. The superscript $a$ represents the scenario where the vertex coupling constants are $\lambda \simeq g \simeq f \equiv 10^{-3}$ and the superscript $b$ represents the scenario where the vertex coupling constants are $\lambda \simeq g \simeq f \equiv 10^{-2}$.}

\begin{ruledtabular}
\begin{tabular}{l|ccccc}                                         
\diagbox{Parameters}{Limits}&KM \cite{takita1986search}&Frejus \cite{berger1990search}&SD-2 \cite{chung2002search}&Super-K \cite{abe2015search} &SNO \cite{aharmim2017search}\\\hline 
Triplet mass $M_{\Delta}$ (TeV)$^a$ &2.18&2.21&2.22&2.35&2.10\\
Neutrino mass $m_\nu$ (eV)$^a$&0.076&0.084&0.085&0.120&0.061\\
VEV $v_{\Delta}$ (eV)$^a$&53.7&59.2&60.0&84.9&43.4\\ \hline
Triplet mass $M_{\Delta}$ (TeV)$^b$ &8.12&8.25&8.27&8.76&7.84\\
Neutrino mass $m_\nu$ (eV)$^b$&0.076&0.084&0.085&0.120&0.061\\
VEV $v_{\Delta}$ (eV)$^b$&5.4&5.9&6.0&8.5&4.3\\ 
\end{tabular}
\end{ruledtabular}
\label{tab3}
\end{table*}

Considering neutrino flavor oscillations, the expected number of the $\nu n \rightarrow \bar{\nu} \bar{n}$ reaction events induced by solar neutrino fluxes can be evaluated as follows:
\begin{eqnarray}
\begin{split}
S 
  =& \epsilon N_n T_n \Bigl[ \sum_{\alpha} p_{\alpha}  \int F_\alpha(E_\nu) \sigma_b(E_\nu) d E_{\nu}\\
   & + \sum_{\beta} p_{\beta}  \int F_\beta(E_\nu) \sigma_b(E_\nu) \delta(E_\nu-E_\beta) d E_{\nu}\Bigl] \\
  =& \frac{\epsilon N_n T_n G_b^2 |\psi_q (0)|^8}{2 \pi} \Bigl[\sum_{\alpha} p_{\alpha}  \int F_\alpha(E_\nu) E_\nu^2 d E_{\nu}\\
   & + \sum_{\beta} p_{\beta}  \int F_\beta(E_\nu) E_\nu^2 \delta(E_\nu-E_\beta) d E_{\nu}\Bigr] \\
\equiv & \frac{1}{2 \pi} \epsilon N_n T_n G_b^2 |\psi_q (0)|^8  \Phi_\nu
\end{split}
\label{signal}
\end{eqnarray}
where $\epsilon$ is the detection efficiency. The index $\alpha$ refers to continuum neutrino sources such as $pp$, $^{13}$N, $^{15}$O, $^{17}$F, $^{8}$B and $hep$. The index $\beta$ refers to mono-energetic neutrino sources such as $^7$Be and $pep$. The factor $p_{\alpha}$ ($p_{\beta}$) stands for the electron neutrino survival probability for the $\alpha$-th ($\beta$-th) component of the solar neutrino sources and their values can be found from Refs. \cite{bellini2014final,abe2016solar}. $F_\alpha$ ($F_\beta$) is the $\alpha$-th ($\beta$-th) component of the solar neutrino fluxes at the Earth, where the experiments are carried out. $T_n$ is the time of data taking. $N_n$ is the number of neutron targets. Such information is summarized in Tab. \ref{tabone} and Tab. \ref{tabtwo}.

In this work, the predicted solar neutrino fluxes from the B16 Standard Solar Model (B16-GS98) \cite{vinyoles2017new} are used for all the solar neutrino sources except for the $^7$Be neutrinos. The $^7$Be neutrinos have two mono-energetic lines with the energy of 0.862 MeV and 0.384 MeV respectively \cite{bellini2014final,miramonti2018solar} and the corresponding fluxes are taken from Ref. \cite{bellini2014final}. The third mono-energetic neutrino source comes from the $pep$ reaction with the energy of $1.44$ MeV \cite{bellini2012first}.

We employ the Bayesian statistical method \cite{helene1983upper,gregory2005bayesian} to evaluate the true number of the $n$-$\bar{n}$ oscillation events. The probability for obtaining $S_0$ candidates can be written as \cite{helene1983upper,gregory2005bayesian}:
\begin{equation}
\begin{split}
P(S_1|S_{0}) &= \frac{1}{N_c} \int \frac{e^{-(S_1 + B_1)} (S_{1}+B_{1})^{S_{0}}}{S_{0}!} g(B_1,B_{0}) dB_1
\end{split}
\label{bayes}
\end{equation}
where $N_c$ is the normalization constant. $S_{1}$ is the true number of events. $B_1$ is the number of background events and $B_0$ is the expected number of background events. $g(B_{1},B_0)$ is the background prior probability density function, which is assumed to be the standard normal distribution. The limit on the true number of events at the $90\%$ confidence level (C.L.) can be determined by the following expression,
\begin{equation}
\begin{split}
\int_0^{S_{max}} P(S_{1}|S_{0}) dS_{1} &= 90\%
\end{split}
\label{confidence}
\end{equation}
In Sec. \ref{section4}, we will show that $\nu n \rightarrow \bar{\nu} \bar{n}$ reactions are unavoidable background noises in search for $n$-$\bar{n}$ oscillations. It is, therefore, reasonable to assume that some of the reported $n$-$\bar{n}$ oscillation candidates are actually contributed from $\nu n \rightarrow \bar{\nu} \bar{n}$ reactions produced by low-energy solar neutrinos with the noise-to-signal ratio $\eta$ ($\eta \in [0,1]$). Using Eq. (\ref{signal}), the derived upper limits on the effective coupling constant $G_b$ for the $\nu n \rightarrow \bar{\nu} \bar{n}$ reaction process at the $90\%$ C.L. can be expressed as follows:
\begin{equation}
\begin{split}
G_b \lesssim \sqrt{\frac{2 \pi \eta S_{max}}{\epsilon N_n T_n |\psi_q (0)|^8 \Phi_\nu}}
\end{split}
\label{gilim}
\end{equation}
In order to quantify the noise-to-signal ratio $\eta$, besides violation of the $(B+L)$ symmetry, we need to focus on the following two different scenarios, depending on whether the $(B-L)$ symmetry could be broken: 
\begin{itemize}
\item[] (A) $(B-L)$ is conserved
\item[] (B) $(B-L)$ could be broken
\end{itemize}

\subsection{$(B-L)$ is conserved}
In this case, we assume that $(B+L)$ could be broken while $(B-L)$ is unbroken, and thus $\nu n \rightarrow \bar{\nu} \bar{n}$ reactions are allowed while $n$-$\bar{n}$ oscillations are forbidden. As it will be explained in Sec. \ref{section4}, the present detectors are unable to distinguish a $\nu n \rightarrow \bar{\nu} \bar{n}$ reaction event from a $n$-$\bar{n}$ oscillation event, and the reported $n$-$\bar{n}$ oscillation candidates are all produced by the solar neutrinos in the scattering process: $\nu n \rightarrow \bar{\nu} \bar{n}$, i.e. $\eta = 1$.

In this case, the bounds on the effective coupling constant $G_b$  can be directly evaluated from Eq. (\ref{gilim}). At the quark level, $G_{b}$ can be expressed as follows:
\begin{equation}
G_b \simeq \frac{g^5}{M_{\Delta}^8}
\end{equation}
where $g$ is the vertex coupling constant and we have assumed that all the relevant vertex coupling constants in Eq. (\ref{fgdefinition}) take similar values, i.e. $\lambda \simeq g_{uu} \simeq g_{dd} \simeq f_{\nu \nu} \equiv  g$. For simplicity, we could choose a natural value $g\simeq 1$ for the vertex coupling constants in our calculation. We have also assumed that all the components of the Higgs triplets have the same mass \cite{costa1982higgs}, i.e. $M_{\Delta_{uu}}\simeq M_{\Delta_{dd}} \simeq M_{\Delta_{\nu \nu}} \equiv M_{\Delta}$. As argued in Ref. \cite{mohapatra1983spontaneous,nieves1984analysis}, those relations can always be satisfied by adjusting the vertex coupling strengths and the masses of the Higgs triplets so that they are compatible with the present limit on the stability of nuclei. The constraint on the mass of the Higgs triplets $M_{\Delta}$, which can be interpreted as the energy scale of new physics, takes the following form,
\begin{equation}
M_{\Delta} \gtrsim \Bigl(\frac{g^{10} \epsilon N_n T_n |\psi_q (0)|^8 \Phi_\nu }{2 \pi S_{max}} \Bigr)^{\frac{1}{16}}
\end{equation}
The bounds at the $90\%$ C.L. on the masses of the Higgs triplets $M_{\Delta}$ and the cross sections of the $\nu n \rightarrow \bar{\nu} \bar{n}$ reaction process can be obtained using the results of the searches for $n$-$\bar{n}$ oscillations inside nuclei from various experiments listed in Tab. \ref{tabtwo}. As we will see in Sec. \ref{section4}, the bounds on the cross sections and the event rates of the $\nu n \rightarrow \bar{\nu}\bar{n}$ reaction process are highly non-trivial and thus plotted in Fig. \ref{crosssection} and Fig. \ref{rate} respectively.

In order to illustrate that the present detectors are unable to distinguish if a particular event is a $n$-$\bar{n}$ oscillation event or a $\nu n \rightarrow \bar{\nu} \bar{n}$ reaction event, we characterize the signal contribution from $\nu n \rightarrow \bar{\nu} \bar{n}$ reactions quantitatively in terms of the signal fraction $\chi_i$, which is defined as
\begin{equation}
\chi_i \equiv \frac{S_i}{\sum S_i}
\end{equation}
where $S_i$ is the number of the $\nu n \rightarrow \bar{\nu} \bar{n}$ reaction events contributed from the $i$-th component of the solar neutrino sources and the sum runs over all the solar neutrino sources. The calculated signal fractions from various solar neutrino sources are presented in Tab. \ref{tabone} and Fig. \ref{rate}.

\subsection{$(B-L)$ could be broken \label{sect3subB}}

In this case, we assume that both $(B+L)$ and $(B-L)$ could be broken, and thus both $\nu n \rightarrow \bar{\nu} \bar{n}$ reactions and $n$-$\bar{n}$ oscillations are allowed. At the quark level, the $n$-$\bar{n}$ oscillation process can be described by a dimension-9 operator while the $\nu n\rightarrow\bar{\nu} \bar{n}$ reaction process can be described by a dimension-12 operator. It is expected that the interaction rate for the $\nu n\rightarrow\bar{\nu} \bar{n}$ reaction process is much smaller than that for the $n$-$\bar{n}$ oscillation process. The noise-to-signal ratio arising from $\nu n\rightarrow\bar{\nu}\bar{n}$ reactions can be evaluated as follows \cite{gavela2016analysis,buchalla2016comment,manohar2018introduction}:
\begin{equation}
\begin{split}
\eta  \simeq  \frac{G_b |\psi_q (0)|^4 P_{\nu}(0)}{G_a |\psi_q (0)|^4} 
\end{split}
\label{eqfraction}
\end{equation}
where the parameters $G_{a}$ and $G_{b}$ represent the coupling constants of the $n$-$\bar{n}$ oscillation process and the $\nu n\rightarrow\bar{\nu}\bar{n}$ reaction process respectively. 
The parameter $P(0) \equiv d_\nu |\psi_\nu(0)|^2$ is the number density of solar neutrinos at the origin where the interaction occurs. The dimensionless parameter $d_\nu$ is the total number of neutrinos inside a neutron and can be estimated very roughly as follows:
\begin{equation}  
d_{\nu} \simeq \frac{4 \pi r_n^3 F_{tot}}{3 v_r}
\end{equation}
Here, $F_{tot}$ is the total flux of solar neutrinos, $r_n \simeq 0.86$ fm \cite{tanabashi2018review} is the neutron radius. The neutrino speed $v_r$ can be replaced with the speed of light, because neutrinos travel at a speed very close to the speed of light \cite{baumann2019first,laha2019constraints}. The parameter $|\psi_\nu(0)|^2$ is the probability density of finding a solar neutrino at the origin. In what follows, we will illustrate that it can be reasonably assumed to be $(\alpha_w m_\nu)^3/\pi$, where $\alpha_w \simeq 0.034$ \cite{tegmark2006dimensionless} is the weak interaction strength. First of all, from Eq. (\ref{eqfraction}), it is easy to see that the parameter $|\psi_\nu(0)|^2$ exhibits the cubic power dependence on neutrino (Lorentz-invariant) mass $m_\nu$ or neutrino energy $E_\nu$, simply because the noise-to-signal ratio $\eta$, which is proportional to the number of the events resulting from $\nu n\rightarrow\bar{\nu}\bar{n}$ reactions, should be a dimensionless constant. Moreover, it is required by Lorentz invariance that the only possible choice for the parameter $|\psi_\nu(0)|^2$ is $m_\nu^3$, rather than $E_\nu^3$. Secondly, a plane-wave description of neutrino faces the problem that the probability of finding it is the same at any point of the whole space and thus leads to an ill-defined parameter $|\psi_\nu(0)|^2$. To solve the problem, a Gaussian wave packet approach has been widely employed to model the neutrino production, interaction, and detection processes in both  non-relativistic and relativistic regimes \cite{blasone2003neutrino,korenblit2017interpolating,daya2017study,akhmedov2019quantum}. Nevertheless, such an approach also has its own problems, one of which, for example, is the difficulty in guessing the form and in quantifying the size of the wave packet \cite{blasone2003neutrino,korenblit2017interpolating,daya2017study,akhmedov2019quantum}. In this work, we assume that the wave function of a solar neutrino can be modeled by a wave packet, and its size can be determined by the interaction between quarks and neutrinos. In the SM, neutrinos only interact with quarks via weak interactions, the strength of which can be characterized by the weak interaction strength $\alpha_w$ \cite{tegmark2006dimensionless}. A greater $\alpha_w$ causes the wave packet to be more contracted on the origin, while a smaller $\alpha_w$ causes the wave packet to be more diffuse. Therefore, it is reasonable to assume that the probability density of finding the neutrino at the origin obeys a power law dependence on $\alpha_w$. Finally, the expression can be determined by comparing it with the corresponding probability density of finding an electron at the origin in the case of $H$-$\bar{H}$ oscillations  \cite{costa1982higgs,nieves1984analysis,mohapatra1983higgs,mohapatra1983spontaneous}. Very roughly, we obtain the following expression, 
\begin{equation}
\begin{split}
P_{\nu}(0)&\equiv d_{\nu} |\psi_{\nu}(0)|^2\\
          &\simeq \frac{4 r_n^3 F_{tot} (\alpha_w m_{\nu})^3}{3 v_r }.
\end{split}
\label{P0}
\end{equation}
Although we have explained that $(\alpha_w m_\nu)^3/\pi$ is a reasonable approximation to $|\psi_{\nu}(0)|^2$ with the help of the wave-packet assumption \cite{blasone2003neutrino,korenblit2017interpolating,daya2017study,akhmedov2019quantum} and the Lorentz invariance requirement, as a matter of fact, it can be obtained by a direct replacement of electron mass ($m_e$) and electromagnetic fine structure constant ($\alpha$) with the neutrino mass ($m_{\nu}$) and the weak interaction strength ($\alpha_{w}$) respectively from the relevant expression used for $H$-$\bar{H}$ oscillations in Refs. \cite{costa1982higgs,nieves1984analysis,mohapatra1983higgs,mohapatra1983spontaneous}.

The vacuum expectation value of the the $\Delta_{\nu \nu}$ field is defined as $\langle \Delta_{\nu \nu} \rangle \equiv v_{\Delta}/\sqrt{2}$. A nonzero $v_{\Delta}$ breaks the $(B-L)$ symmetry spontaneously and can be related to the neutrino mass by the following expression \cite{mohapatra1983spontaneous,dev2018doubly}
\begin{equation}
m_\nu = \sqrt{2}f_{\nu \nu}v_{\Delta}
\label{mass}
\end{equation}

The $n$-$\bar{n}$ oscillation process depicted in Fig. \ref{figab} (a) has been intensively studied \cite{mohapatra1980local,mohapatra1983higgs,chacko1999supersymmetric,dutta2006observable,mohapatra2009neutron,babu2009neutrino,patraa2014post} and the corresponding coupling constant can be given by \cite{mohapatra1982hydrogen,costa1982higgs,mohapatra1983higgs,dutta2006observable,mohapatra2009neutron,babu2012coupling,babu2009neutrino,patraa2014post}
\begin{equation}
G_a \simeq \frac{g_{uu} g_{dd}^2 \lambda v_{\Delta}}{M_{\Delta_{uu}}^2 M_{\Delta_{dd}}^4}
\end{equation}

Throughout this work, we assume that neutrinos only have Majorana masses. However, it would be problematic, if one assumes that neutrino acquires a Majorana mass directly from the spontaneous breaking of the $(B-L)$ symmetry. To begin with, if the $(B-L)$ symmetry breaks down spontaneously at the energies above the electroweak scale, then in order to generate tiny neutrino masses the Yukawa coupling constant $f_{\nu\nu}$ should be much smaller than the ones in the quark sector, which is considered to be highly unnatural. Furthermore, the vacuum expectation value $v_{\Delta}$ contributes differently to the masses of the $W$ and $Z$ bosons after the electroweak symmetry breaking, and then it affects the $\rho$-parameter \cite{arhrib2011higgs,kanemura2012radiative,dev2013125} in the following way
\begin{equation}
\rho \simeq \frac{v_{H}^2 + 2 v_{\Delta}^2}{v_{H}^2 + 4 v_{\Delta}^2}
\end{equation}
where $v_{H}$ is the vacuum expectation value of the $SU(2)_L$ Higgs doublet and satisfies the relation: $v_H^2 + v_{\Delta}^2 \simeq$ $(246$ GeV$)^2$ \cite{li2018type,dey2019inverse}. The $\rho$-parameter describes the relative coupling strength between the Higgs bosons and the gauge bosons, and can be precisely determined from experiments. The upper bounds on $v_{\Delta}$ imposed by precision electroweak data, such as measurements on the $\rho$-parameter, are approximately at the order of 1 GeV \cite{del2008electroweak,akeroyd2009lepton,kanemura2012radiative,dev2013125,chabab2016naturalness,dev2017naturalness,pan2019triply,padhan2019probing,antusch2019low}. The lower bounds on $v_{\Delta}$, arising from the cosmological observations and the measurements of the lepton flavor violating (LFV) processes (see e.g. Ref. \cite{baldini2016search}), are approximately at the order of 1 eV \cite{akeroyd2009lepton,chen2011type,melfo2012type,dev2017naturalness,de2019implementing,pan2019triply}. In this work, we therefore reasonably require that the vacuum expectation value $v_{\Delta}$ satisfies the condition $1$ eV $ \lesssim v_{\Delta} \lesssim$ $1$ GeV \cite{del2008electroweak,akeroyd2009lepton,chen2011type,melfo2012type,kanemura2012radiative,dev2013125,chabab2016naturalness,dev2017naturalness,de2019implementing,pan2019triply,antusch2019low,padhan2019probing}. Finally, a massless particle called Majoron \cite{chikashige1981there,schechter1982neutrino} can be produced from the spontaneous breaking of the $(B-L)$ symmetry but it has been ruled out by the precise measurements of $Z$ boson decay \cite{ma2017pseudo,tanabashi2018review}.

The above problems may be solved by the type-\RomanNumeralCaps{2} seesaw mechanism \cite{cheng1980neutrino,mohapatra1980neutrino,schechter1980neutrino,lazarides1981proton}, which employs the following potential in describing the interactions between the Higgs doublet ($\Phi$) and triplet ($\Delta_l$) \cite{ma2001phenomenology,chun2003testing,akeroyd2009lepton,arhrib2011higgs,arhrib2012higgs,dev2013125,du2019type}:
\begin{equation}
\begin{split}
V(\Phi, \Delta_l)  = & -M_H^2 \Phi^{\dagger}\Phi + \frac{\lambda_0}{4} \bigl(\Phi^{\dagger}\Phi\bigr)^2 + M_\Delta^2 \text{Tr}\bigl(\Delta_l^{\dagger} \Delta_l \bigr) \\
                & + \lambda_1 \bigl(\Phi^{\dagger}\Phi \bigr)\text{Tr}\bigl(\Delta_l^{\dagger} \Delta_l \bigr) + \lambda_2 \bigl[ \text{Tr}\bigl(\Delta_l^{\dagger} \Delta_l \bigr)\bigr]^2 \\
                &+ \lambda_3 \text{Tr}\bigl[\bigl(\Delta_l^{\dagger} \Delta_l \bigr)^2\bigr]+ \lambda_4 \Phi^{\dagger}\Delta_l \Delta_l^{\dagger} \Phi\\
                & + \bigl[\mu \Phi^{T}i \sigma_2 \Delta_l^{\dagger} \Phi + \text{H.c.}\bigr]
\end{split}
\label{seesaw}
\end{equation}
where $M_H$ is the mass of the Higgs doublet $\Phi$ and $M_\Delta$ are the masses of the newly added Higgs triplet $\Delta_l$ defined in Sec. \ref{section2}. Here, we assume that all the components of the Higgs triplets have the same mass, i.e. $M_{\Delta_{ee}}\simeq M_{\Delta_{\nu e}} \simeq M_{\Delta_{\nu \nu}} \equiv M_{\Delta}$. The $\mu$-term in Eq. (\ref{seesaw}) eliminates Majoron and violates the lepton number by two units ($|\Delta L| = 2$) \cite{chen2011type}. In the type-\RomanNumeralCaps{2} seesaw mechanism, the following vacuum expectation value $v_{\Delta}$ can be obtained by minimizing the potential $V(\Phi, \Delta_l)$ \cite{chen2011type,melfo2012type,dev2017naturalness,li2018type}
\begin{equation}
v_{\Delta} \simeq \frac{\mu v_H^2}{\sqrt{2} M_\Delta^2}
\label{vev}
\end{equation}
Actually, the vacuum expectation value $v_{\Delta}$ not only can be given by Eq. (\ref{vev}) but also can be given by Eq. (\ref{mass}). In this work, we employ Eq. (\ref{mass}) to evaluate $v_{\Delta}$, but we can always adjust the parameter $\mu$ \cite{chen2019radiatively}, so that the value of $v_{\Delta}$ given by Eq. (\ref{vev}) also satisfies the bounds $1$ eV $ \lesssim v_{\Delta} \lesssim$ $1$ GeV  \cite{del2008electroweak,akeroyd2009lepton,chen2011type,melfo2012type,kanemura2012radiative,dev2013125,chabab2016naturalness,dev2017naturalness,de2019implementing,pan2019triply,antusch2019low,padhan2019probing}.

Similar to $H$-$\bar{H}$ oscillations \cite{mohapatra1982hydrogen,costa1982higgs,arnellos1982hydrogen,mohapatra1983higgs,deo1984hydrogen,mohapatra1983spontaneous,nieves1984analysis,alberico1985double}, the coupling constant for the $\nu n\rightarrow\bar{\nu} \bar{n}$ reaction process depicted in Fig. \ref{figab} (b) can be written as
\begin{equation}
G_b \simeq \frac{g_{uu} g_{dd}^2 f_{\nu \nu} \lambda}{M_{\Delta_{uu}}^2 M_{\Delta_{dd}}^4 M_{\Delta_{\nu \nu}}^2}
\end{equation}

Using the above equations, the noise-to-signal ratio arising from $\nu n \rightarrow\bar{\nu} \bar{n}$ reactions can be written as
\begin{equation}
\eta \simeq \frac{4 r_n^3 F_{tot} f_{\nu \nu} \alpha_w^3 m_{\nu}^3}{3 v_r v_{\Delta} M_{\Delta_{\nu \nu}}^2}
\end{equation}
It is reasonable to assume that the vertex coupling constants take similar values, i.e. $g_{uu} \simeq g_{dd} \equiv g$ for the quark sector and $f_{ee} \simeq f_{\nu\nu} \equiv f$ for the lepton sector. Moreover, considering the requirement of naturalness, throughout this work, if not otherwise mentioned, we assume that the coupling constants in the lepton sector should be similar to the ones in the quark sector, as well as to the ones in the gauge sector, i.e. $\lambda \simeq g \simeq f$. Similarly, it is also reasonable to assume that, the Higgs triplets, namely the diquark and dilepton fields, which are responsible for $n$-$\bar{n}$ oscillations and $\nu n \rightarrow \bar{\nu} \bar{n}$ reactions, have the same mass $M_{\Delta}$ \cite{costa1982higgs}, i.e. $M_{\Delta_{uu}}\simeq M_{\Delta_{dd}} \simeq M_{\Delta_{\nu \nu}} \equiv M_{\Delta}$. Again, the parameter $M_{\Delta}$ represents the mass of the Higgs triplet bosons and can also be interpreted as the energy scale of new physics. In this work, we employ one of the popular ways of explaining the small but nonzero neutrino mass by assuming that neutrino only has a Majorana mass, which is generated within the simplest type-\RomanNumeralCaps{2} seesaw framework \cite{dev2018doubly}. Under such assumptions, the lower bound on the mass of the Higgs triplets $M_{\Delta}$ arising from the results of the searches for $n$-$\bar{n}$ oscillations presented in Tab. \ref{tabtwo} can be written as
\begin{equation}
M_{\Delta} \gtrsim \Bigl( \frac{3 g^8 v_r \epsilon N_n T_n \Phi_\nu |\psi_q(0)|^8}{8 \sqrt{2} \pi S_{max} r_n^3 F_{tot} \alpha_w^3 m_\nu^2}\Bigr)^{\frac{1}{14}}
\label{lq1}
\end{equation}

On the other hand, the direct search from the ILL experiment shows that the $n$-$\bar{n}$ oscillation time satisfies the bound $\tau_{n-\bar{n}} \gtrsim 0.86 \times 10^8$ s \cite{baldo1994new} or, equivalently, $\delta m \equiv 1/\tau_{n-\bar{n}} \lesssim 7.65 \times 10^{-33}$ GeV ($\hbar \equiv 1$). Here, the parameter $\delta m$ can also be written as
\begin{equation}
\delta m \equiv G_a |\psi_q(0)|^4
\end{equation}
The corresponding bound on $M_{\Delta}$ arising from the direct search can be expressed as a function of the parameter $\delta m$:
\begin{equation}
M_{\Delta} \gtrsim \Bigl( \frac{g^4 m_\nu |\psi_q(0)|^4}{\sqrt{2} f \delta m }\Bigr)^{\frac{1}{6}}
\label{lq2}
\end{equation}
The bound on the $n$-$\bar{n}$ oscillation time can be obtained from Eq. (\ref{lq2}):
\begin{equation}                                                               
\tau_{n-\bar{n}} \gtrsim \frac{\sqrt{2}fM_{\Delta}^6}{g^4 m_\nu |\psi_q(0)|^4}
\label{tnn}
\end{equation}
Obviously, the $n$-$\bar{n}$ oscillation time is sensitive to the vertex coupling constants and the masses of Higgs triplet bosons.

In addition to the condition given by Eq. (\ref{lq2}), an additional condition given by Eq. (\ref{lq1}) is obtained from the measurements of $n$-$\bar{n}$ oscillations inside nuclei. Eq. (\ref{lq1}) and Eq. (\ref{lq2}) depend on the neutrino mass $m_\nu$ in a different way but we could adjust the parameters, such as $g$, $f$, and $\delta m$, so that both of them can be incorporated into the analysis in a compatible way. Meanwhile, the bounds on the sum of neutrino masses have been reported in various cosmological scenarios \cite{choudhury2018updated,aghanim2018planck,mahony2019target}. Recently, an upper bound of $0.12$ eV on the sum of neutrino masses has been established at a $95\%$ C.L. by cosmological measurements \cite{aghanim2018planck}. Throughout our analysis, we assume that the neutrino mass satisfies the condition: $m_\nu \lesssim \sum m_{\nu} \lesssim 0.12$ eV \cite{aghanim2018planck}, where the sum runs over the three mass eigenstates. The bounds on neutrino masses impose further constraints on the parameter space. As we will see later, since the Super-K experiment provides the most stringent bounds, in practice we require that the constrained curve arising from the Super-K experiment \cite{abe2015search} intersects the constrained curve arising from Eq. (\ref{lq2}) at the neutrino mass of around $0.12$ eV \cite{aghanim2018planck}. In order to satisfy this requirement, the parameter $\delta m$ in Eq. (\ref{lq2}) has to be adjusted. In other words, Eq. (\ref{lq2}) could be used to predict the $n$-$\bar{n}$ oscillation time. In Fig. \ref{constraints} and Fig. \ref{constraints2}, the dashed curves represent the constraints of Eq. (\ref{lq1}) arising from the results of the searches for $n$-$\bar{n}$ oscillations inside nuclei, while the solid curve represents the theoretical prediction (TP) of Eq. (\ref{lq2}) on $n$-$\bar{n}$ oscillations. As it can be seen, the dashed and the solid curves intersect at the allowed neutrino masses, i.e. $m_\nu \lesssim$ $0.12$ eV \cite{aghanim2018planck}. In the vicinity of the intersections, a smaller neutrino mass than the one at the intersection is forbidden by Eq. (\ref{lq1}) while a greater neutrino mass is forbidden by Eq. (\ref{lq2}). The intersections between the dashed and the solid curves provide the minimum possible mass of the Higgs triplets. As mentioned early, the naturalness consideration requires that the vertex coupling constants in the lepton sector should be similar to the ones in the quark sector, as well as to the ones in the gauge sector, i.e. $\lambda \simeq g \simeq f$. Furthermore, the bounds on the vacuum expectation value, $1$ eV $ \lesssim v_{\Delta} \lesssim$ $1$ GeV \cite{del2008electroweak,akeroyd2009lepton,chen2011type,melfo2012type,kanemura2012radiative,dev2013125,chabab2016naturalness,dev2017naturalness,de2019implementing,pan2019triply,antusch2019low,padhan2019probing}, should also be taken into account. The above conditions severely constrain the parameter space. The acceptable values of the parameters on the two processes are presented in Tab. \ref{tab3}, where the vacuum expectation value $v_{\Delta}$ is evaluated from Eq. (\ref{mass}). The derived bounds on the mass of the Higgs triplet $M_{\Delta}$ as a function of the neutrino mass $m_{\nu}$ are plotted in Fig. \ref{constraints} and Fig. \ref{constraints2} with the allowed vertex coupling constants $10^{-3}$ and $10^{-2}$ respectively.

\section{Result and discussion} \label{section4}

In the following discussions, we will focus on the two different scenarios, depending on whether the $(B-L)$ symmetry could be broken. We will first illustrate that in both scenarios due to the low-energy thresholds for neutrino detection, the present experiments are unable to distinguish if a particular event is a $n$-$\bar{n}$ oscillation event or a $\nu n \rightarrow \bar{\nu} \bar{n}$ reaction event. Moreover, we will also investigate the interplay of various conditions on the parameter space and their observable consequences.

\subsection{$(B-L)$ is conserved}

\begin{figure}[t] 
\centering
\includegraphics[scale=0.90,width=0.98\linewidth]{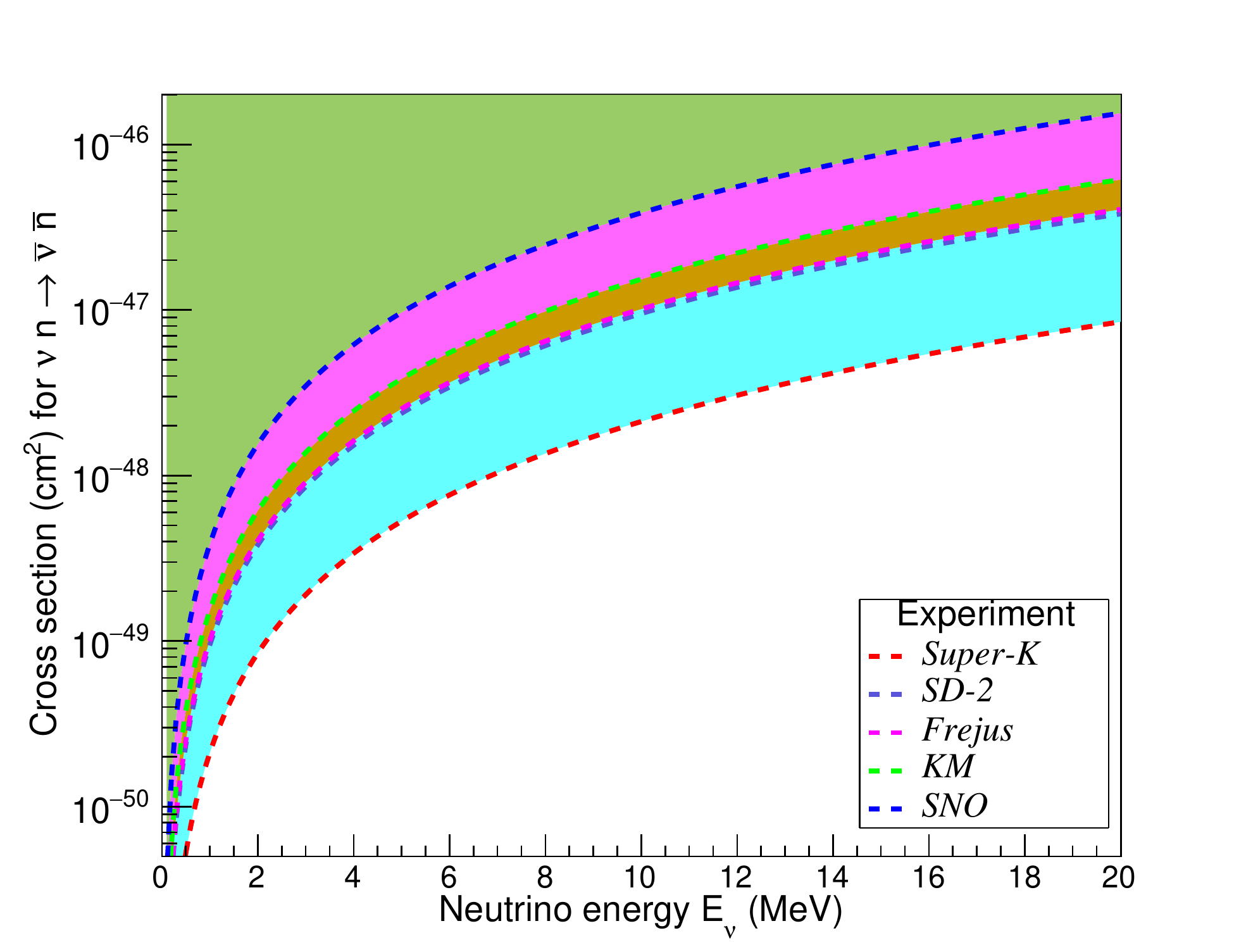}
\caption{Upper bounds at the $90\%$ C.L. on the cross sections of the $\nu n \rightarrow \bar{\nu} \bar{n}$ reaction process imposed by the $n$-$\bar{n}$ oscillation experiments in the range $m_\nu \ll E_\nu \ll m_n$ in scenario A. (Color online)
}
\label{crosssection}
\end{figure}

\begin{figure}[t]    
\centering
\includegraphics[scale=0.90,width=0.98\linewidth]{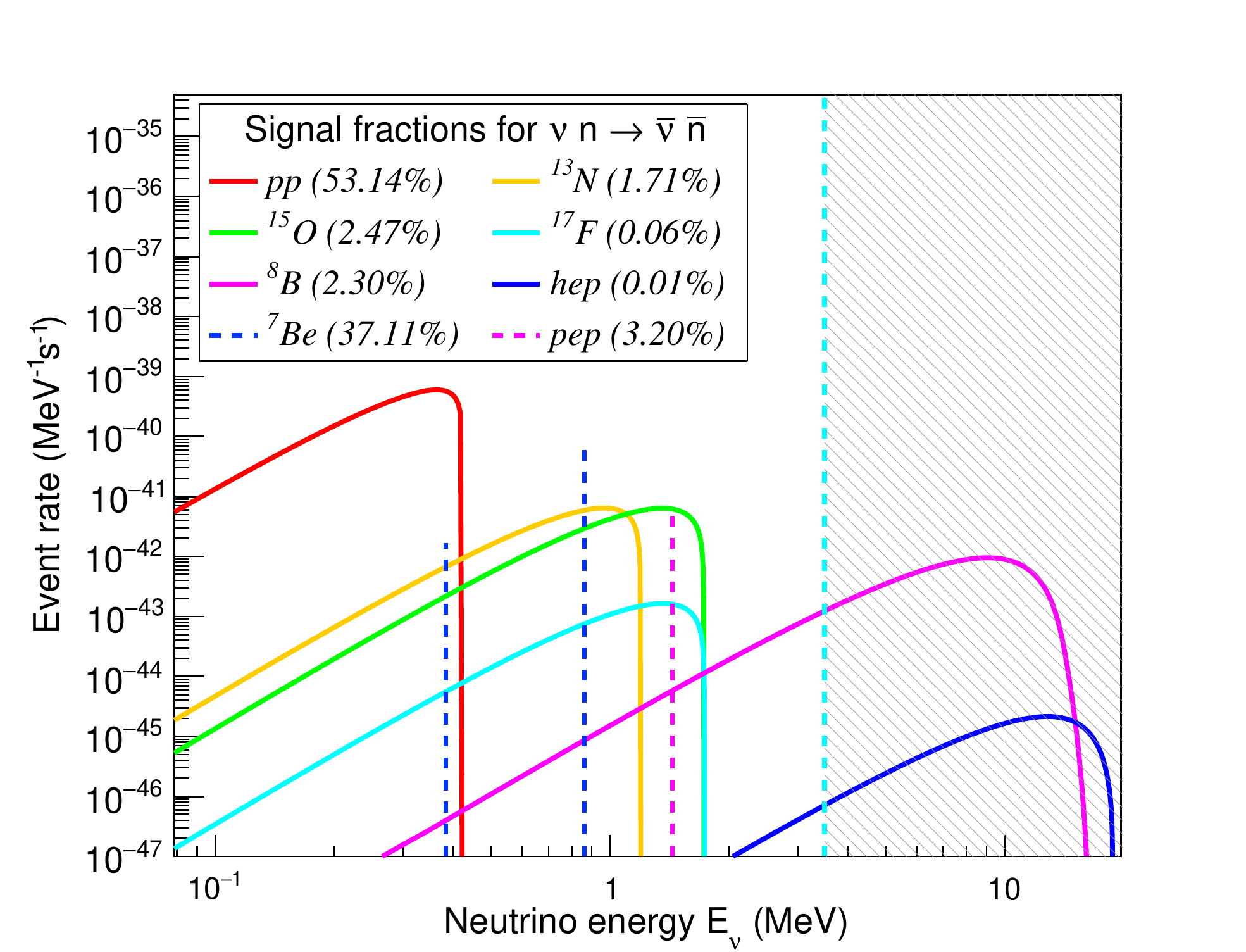}
\caption{Upper bounds at the $90\%$ C.L. imposed by the Super-K data on the event rates of the $\nu n \rightarrow \bar{\nu} \bar{n}$ reaction process produced by various solar neutrino sources in scenario A.  
(Color online)}
\label{rate}
\end{figure}

Tab. \ref{tabone} summarizes the fluxes, survival probabilities, and the corresponding signal fractions for (electron-type) solar neutrinos from various sources. In this table, the probability densities of the solar neutrino fluxes are taken from Refs. \cite{Bahcall1987jc,bahcall1996standard,bahcall1997gallium,bahcallweb} and the fluxes are normalized according to Refs. \cite{vinyoles2017new,bellini2014final}. It is remarkable that the low-energy $pp$ neutrinos make up more than $90\%$ of the total solar neutrino fluxes \cite{altmann2005complete,hsieh2019discovery}, but such neutrinos have very low energies, which only cover the range below $420$ keV \cite{kirsten1999solar,bellini2014neutrinos,hsieh2019discovery}. Our calculation shows that the $pp$ neutrinos make the largest contribution ($\sim 53.14\%$) to $\nu n \rightarrow \bar{\nu} \bar{n}$ reactions because of its relatively higher intensity. Then, it is followed by the $^7$Be neutrinos with the signal fraction of around $\sim 37.11\%$. Therefore, $\nu n\rightarrow \bar{\nu}\bar{n}$ reactions are dominated by the $pp$ and $^7$Be solar neutrinos with the summed signal fraction of around $\sim 90.25\%$. However, the $pp$ and $^7$Be solar neutrinos have an energy range that is not accessible to the detectors listed in Tab. \ref{tabtwo}, because such detectors can only detect neutrinos above an energy threshold of around $3.5$ MeV \cite{giganti2018neutrino}. The calculation also shows that more than $92.07\%$ of the contribution to $\nu n \rightarrow \bar{\nu} \bar{n}$ reactions comes from solar neutrinos with energies lower than $1.0$ MeV. Particularly, the contribution fraction within energy range from $0.2$ MeV to $1.0$ MeV is as high as $88.52\%$. At such energies, the outgoing antineutrinos are completely invisible to the detectors under discussion. Therefore, the detectors listed in Tab. \ref{tabtwo} are unable to distinguish between a $n$-$\bar{n}$ oscillation event and a $\nu n \rightarrow \bar{\nu} \bar{n}$ reaction event.

In this case, the derived upper bounds at the $90\%$ C.L. on the cross section of the $\nu n \rightarrow \bar{\nu} \bar{n}$ reaction process is shown in Fig. \ref{crosssection}, where the shaded regions are excluded by the $n$-$\bar{n}$ oscillation experiments. As it can be seen, the most stringent constraint on the cross section is imposed by the Super-K experiment. Fig. \ref{rate} shows the derived bounds on the event rate of $\nu n \rightarrow \bar{\nu} \bar{n}$ reactions imposed by the Super-K data, where the shaded region is visible to the detectors under discussion. For a natural value of the vertex coupling strength ($\lambda \simeq g \simeq f \equiv 1$), the derived bounds on the masses of the Higgs triplets $M_{\Delta}$ imposed by the Super-K experiment is roughly $\sim 3$ GeV. Although such bounds on the masses of the Higgs triplets, which are model dependent, seem not very useful, the derived bounds on the cross sections of the $\nu n \rightarrow \bar{\nu} \bar{n}$ reaction process are highly non-trivial. For example, the derived bound on the cross section at the average neutrino energy from the Super-K data is around $6.0 \times 10^{-51}$ cm$^2$, which is much smaller than the ones given by the typical electroweak and some non-standard neutrino-nucleon interactions \cite{papoulias2015standard,papoulias2019recent,sharma2018status}. A reasonable interpretation of such results requires further phenomenological studies using an appropriate effective model.

In this case, since $(B-L)$ is conserved, all the reported $n$-$\bar{n}$ oscillation candidates are actually produced by solar neutrinos in the scattering process $\nu n \rightarrow \bar{\nu} \bar{n}$, it would then be possible to distinguish a $n$-$\bar{n}$ oscillation event from a $\nu n \rightarrow \bar{\nu} \bar{n}$ reaction event. In order to distinguish such two processes, it is essential to employ detectors with detectable range covering the $pp$ and $^7$Be solar neutrinos. On the contrary, the $^8$B neutrinos, which are relatively more easy to be measured in the Super-K detector \cite{rosso2018introduction}, only contribute a very small fraction ($\sim$ $2.30\%$) to the $\nu n \rightarrow \bar{\nu} \bar{n}$ reaction signal. The future Hyper-Kamiokande (Hyper-K) detector is designed to use 187 kton of water \cite{abe2018hyper}, corresponding to $5.0 \times 10^{34}$ neutrons approximately. The expected event rate of $\nu n \rightarrow \bar{\nu} \bar{n}$ reactions in Hyper-K is around 49 events per year, which is roughly 8 times higher than that in Super-K, thus reducing the impact of backgrounds considerably. Some other experiments such as GALLEX \cite{Anselmann1992um}, SAGE \cite{Abdurashitov1994bc}, LOREX \cite{pavicevic2018lorandite}, and Borexino \cite{bellini2014neutrinos} have been sensitive to low-energy $pp$ neutrinos. These experiments also provide a good opportunity to study $\nu n \rightarrow \bar{\nu} \bar{n}$ reactions and might help distinguish a $n$-$\bar{n}$ oscillation event from a $\nu n \rightarrow \bar{\nu} \bar{n}$ reaction event.

Besides the solar neutrinos, there are a number of other neutrino sources \cite{bahcall2001proceedings,xing2011neutrinos,katz2012high}, each of which has its own spectrum with a particular shape of distribution \cite{katz2012high,halzen2014highest}. Neutrinos from such sources cover a wide range of energies from $10^{-10}$ MeV to $10^{8}$ MeV \cite{becker2008high,ringwald2009prospects,IceCube2018dnn}. According to Eq. (\ref{signal}), different neutrino sources contribute differently to $\nu n \rightarrow \bar{\nu} \bar{n}$ reactions. It is worth mentioning that the cosmic neutrino background has a even higher intensity but only has an average energy of around $10^{-10}$ MeV \cite{ringwald2009prospects} and thus its contribution to $\nu n \rightarrow \bar{\nu} \bar{n}$ reactions is not significant. The supernova neutrinos are predicted to be evenly distributed among the three flavors of particles and antiparticles \cite{snoweb,spurio2014particles}. The summed flux of all neutrino types at the Earth for a supernova at 10 kpc distance is about $10^{12}$ cm$^{-2}$ with an average energy of around $15$ MeV \cite{abbasi2011icecube}. The expected number of events in the future Hyper-K is around $0.04$ per supernova burst, much smaller than that produced by the solar neutrinos. The fluxes of the rest neutrino sources are much smaller than that of the solar neutrinos and because of the limited statistics they also have very little impact on $\nu n \rightarrow \bar{\nu} \bar{n}$ reactions. Unlike the solar neutrinos, the reactor neutrinos are mainly electron antineutrinos. Detecting electron antineutrinos is relatively easier than detecting electron neutrinos. The relevant possible process leading to the instability of nuclei is the $\bar{\nu} n \rightarrow \nu \bar{n}$ reaction process. Although such reaction preserves the $(B+L)$ symmetry, it violates the $(B-L)$ symmetry and thus contradicts our basic assumption in this scenario.

\subsection{$(B-L)$ could be broken}

In this case, both $\nu n \rightarrow \bar{\nu} \bar{n}$ reactions and $n$-$\bar{n}$ oscillations are allowed according to the assumption. Comparing with the $n$-$\bar{n}$ oscillation process, the $\nu n \rightarrow \bar{\nu} \bar{n}$ reaction process can be described by higher-dimensional operators, and thus the effects are strongly suppressed by appropriate powers of energy scale associated with new physics, causing the signal too small to be detectable. Obviously, in this case, the detectors listed in Tab. \ref{tabtwo} are still unable to distinguish a $\nu n\rightarrow \bar{\nu} \bar{n}$ reaction event from a the $n$-$\bar{n}$ oscillation event.

We next explore the interplay of the following conditions on the parameter space for the two processes within the type-\RomanNumeralCaps{2} seesaw framework: (\romannumeral 1) The condition given by Eq. (\ref{lq1}) arising from the results of the searches for $n$-$\bar{n}$ oscillations inside nuclei should be satisfied; (\romannumeral 2) The condition given by Eq. (\ref{lq2}), directly related to the $n$-$\bar{n}$ oscillation time, should be satisfied; (\romannumeral 3) The neutrino mass should at least satisfy the experimental constraint on the sum of the neutrino masses, i.e. $m_\nu \lesssim \sum m_{\nu} \lesssim 0.12$ eV \cite{aghanim2018planck}; (\romannumeral 4) The naturalness criterion of the vertex coupling constants should be fulfilled; (\romannumeral 5) The vacuum expectation values of the Higgs triplet bosons $v_{\Delta}$ should satisfy the bounds: $1$ eV $ \lesssim v_{\Delta} \lesssim$ $1$ GeV \cite{del2008electroweak,akeroyd2009lepton,chen2011type,melfo2012type,kanemura2012radiative,dev2013125,chabab2016naturalness,dev2017naturalness,de2019implementing,pan2019triply,antusch2019low,padhan2019probing}; (\romannumeral 6) The mass of the Higgs triplet bosons should be in the experimentally interesting range at the LHC or future high-energy experiments \cite{mohapatra2008diquark,dev2017naturalness,nomura2018discriminating,nomura2018discriminating,du2019type,de2019doubly}. Therefore, it is expected that if all such requirements are satisfied, the parameter space will be severely constrained.

Specifically, we are interested in the appealing scenario where the mass of the Higgs triplet bosons is in the several TeV range ($1$ TeV $\lesssim M_{\Delta} \lesssim$ $10$ TeV), which is expected to lie within the reach of direct searches at the LHC or future high-energy experiments \cite{mohapatra2008diquark,dev2017naturalness,nomura2018discriminating,nomura2018discriminating,du2019type,de2019doubly}. For simplicity, we have neglected the mass splitting of all the triplet components by assuming $M_{\Delta_{uu}}\simeq M_{\Delta_{dd}} \simeq M_{\Delta_{\nu \nu}} \equiv M_{\Delta}$. The experimental lower bounds on the mass of the doubly-charged Higgs bosons set by the LHC data are approximately in the range from 450 GeV to 870 GeV \cite{atlas2016search,cms2017search,aaboud2018search}. Considering the detectable several TeV scale triplet mass ($1$ TeV $\lesssim M_{\Delta} \lesssim$ $10$ TeV) and the experimental bounds on the neutrino mass ($m_{\nu} \lesssim$ $0.12$ eV \cite{aghanim2018planck}), as well as on the vacuum expectation value ($1$ eV $ \lesssim v_{\Delta} \lesssim$ $1$ GeV \cite{del2008electroweak,akeroyd2009lepton,chen2011type,melfo2012type,kanemura2012radiative,dev2013125,chabab2016naturalness,dev2017naturalness,de2019implementing,pan2019triply,antusch2019low,padhan2019probing}), the parameter scan shows that the vertex coupling constants ($\lambda \simeq g \simeq f$) are roughly restricted in the range from the order of $10^{-3}$ to the order of $10^{-2}$. A greater coupling constant ($f \gtrsim 10^{-1}$) would lead to a too small vacuum expectation value, which does not satisfy the lower bound $v_{\Delta} \gtrsim 1$ eV 
\cite{akeroyd2009lepton,chen2011type,melfo2012type,dev2017naturalness,de2019implementing,pan2019triply}, and it would also give rise to a too large triplet mass ($M_{\Delta} \gtrsim 10$ TeV), which is probably beyond the reach of a direct detection at the LHC. A smaller coupling constant ($f \lesssim 10^{-4}$) would lead to a too small triplet mass ($M_{\Delta} \lesssim 700$ GeV), which, in general, does not satisfy the experimental lower bounds on the mass of the doubly-charged Higgs bosons set by the LHC data \cite{atlas2016search,cms2017search,aaboud2018search}.

\begin{figure}[t] 
\centering
\includegraphics[scale=0.90,width=0.98\linewidth]{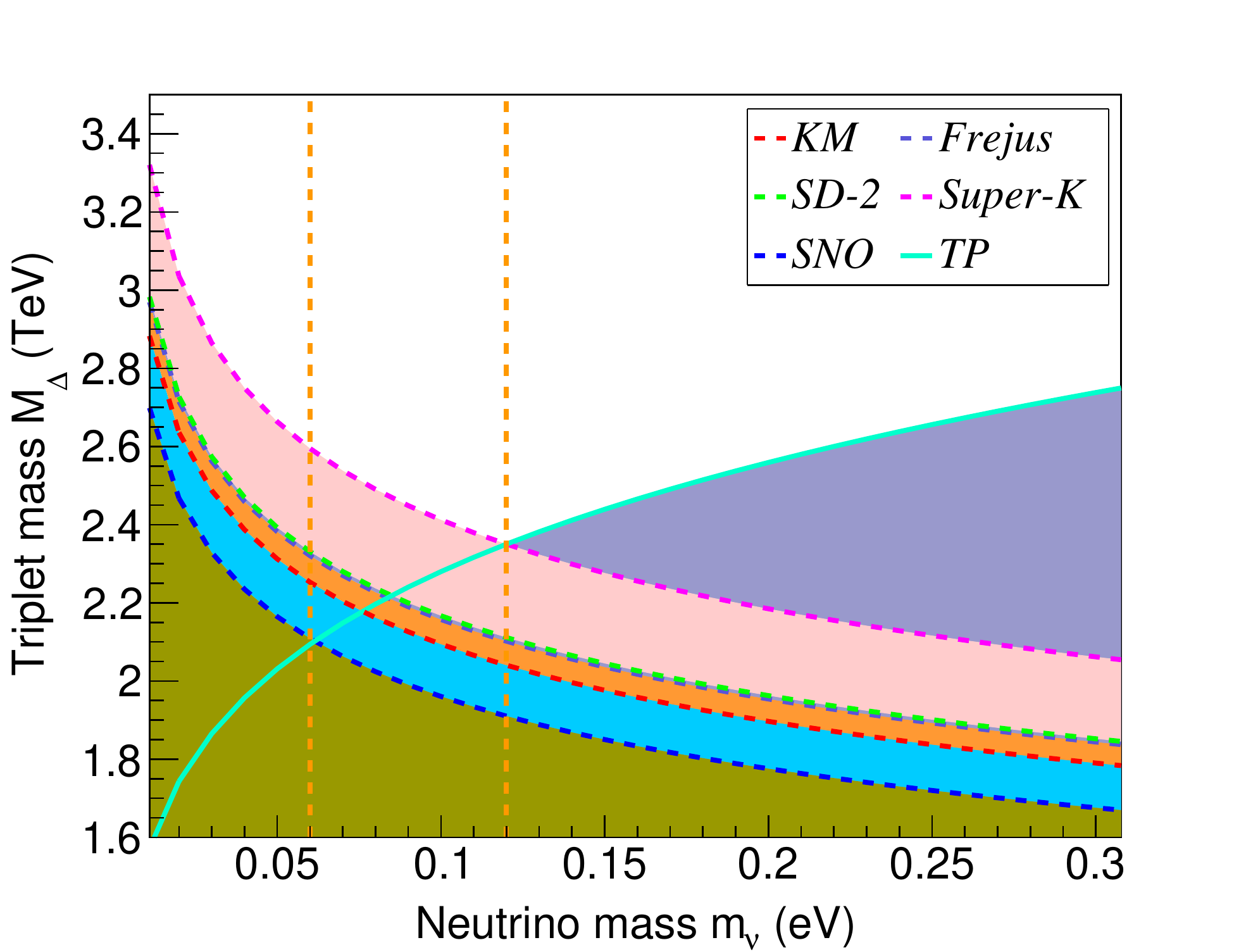}
\caption{Exclusion curves for the masses of the Higgs triplet bosons $M_{\Delta}$ (TeV) as a function of neutrino masses $m_\nu$ (eV) in the scenario where the vertex coupling constant is $10^{-3}$. (Color online)}
\label{constraints}
\end{figure}

\begin{figure}[t] 
\centering
\includegraphics[scale=0.90,width=0.98\linewidth]{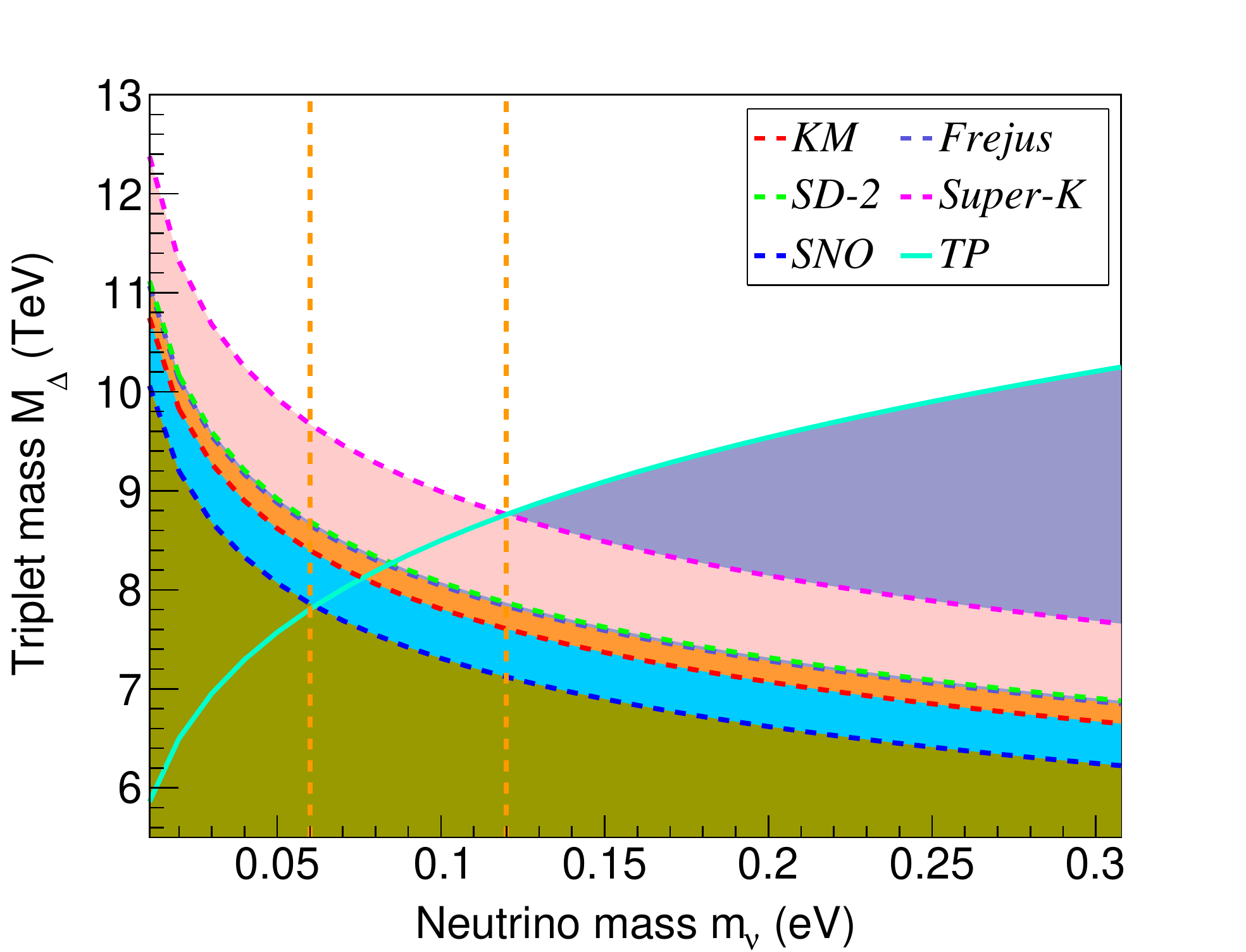}
\caption{Exclusion curves for the masses of the Higgs triplet bosons $M_{\Delta}$ (TeV) as a function of neutrino masses $m_\nu$ (eV) in the scenario where the vertex coupling constant is $10^{-2}$. (Color online)}
\label{constraints2}
\end{figure}

Fig. \ref{constraints} and Fig. \ref{constraints2} show the bounds on the masses of the Higgs triplet bosons $M_{\Delta}$ as a function of neutrino masses in the scenarios where the vertex coupling constants ($\lambda \simeq g \simeq f$) are $10^{-3}$ and $10^{-2}$ respectively. The dashed curves in both plots satisfy the constraints imposed by Eq. (\ref{lq1}). The solid curve in both plots satisfies the constraints imposed by Eq. (\ref{lq2}). The dashed vertical lines represent the experimental lower and upper bounds on the sum of neutrino masses in the cosmological scenario \cite{choudhury2018updated,mahony2019target,aghanim2018planck}.

Tab. \ref{tab3} presents the derived bounds on masses of the Higgs triplet bosons, the neutrino masses ($m_{\nu}$), and the vacuum expectation values ($v_{\Delta}$) using the acceptable values of the vertex coupling constants ($\lambda \simeq g \simeq f$). The parameters in the upper part of Tab. \ref{tab3} correspond to the coupling constant $10^{-3}$. In this case, the bounds on the masses of the Higgs triplet are approximately in the range from $2.10$ TeV to $2.35$ TeV, which can be accessible to a direct detection at the LHC or future high-energy experiments \cite{mohapatra2008diquark,dev2017naturalness,nomura2018discriminating,nomura2018discriminating,du2019type,de2019doubly}. The parameters in the lower part of Tab. \ref{tab3} correspond to the coupling constant $10^{-2}$. In this case, the bounds on the masses of the Higgs triplet are approximately in the range from $7.84$ TeV to $8.76$ TeV, which may still be within the reach of direct searches at the LHC or future high-energy experiments \cite{mohapatra2008diquark,dev2017naturalness,nomura2018discriminating,nomura2018discriminating,du2019type,de2019doubly}. As can be seen from Tab. \ref{tab3}, in both scenarios ($10^{-3}$, $10^{-2}$), the differences in the bounds on the masses of the Higgs triplet from different experiments are less than $1.0$ TeV. This illustrates that the derived bounds on the masses of the Higgs triplet, i.e. the energy scales of new physics, depend weakly on the reported number of candidate events due to the fractional power dependence of Eq. (\ref{lq1}). The existing data from the Super-K experiment provides leading bounds, ruling out the existence of new physics below energy scale of $2.4$ TeV and $8.8$ TeV, depending on the choice of the vertex coupling strengths respectively. For this reason, we will give a special attention to the Super-K experiment in the following discussions.

The $n$-$\bar{n}$ oscillation time can be easily estimated using the acceptable parameters for the Super-K experiment in Tab. \ref{tab3}. If we choose $f \simeq 10^{-3}$, $M_{\Delta} \simeq 2.35$ TeV, $m_{\nu} \simeq 0.12$ eV \cite{aghanim2018planck}, we get the $n$-$\bar{n}$ oscillation time $\tau_{n\bar{n}} \gtrsim 6.3 \times 10^{18}$ s. Similarly, if we choose $f \simeq 10^{-2}$, $M_{\Delta} \simeq 8.76$ TeV, $m_{\nu} \simeq 0.12$ eV \cite{aghanim2018planck}, we get the $n$-$\bar{n}$ oscillation time $\tau_{n\bar{n}} \gtrsim 1.7 \times 10^{19}$ s. This illustrates that the $n$-$\bar{n}$ oscillation effects in both two cases are probably beyond the reach of the present experiments. Therefore, if we assume that all the requirements listed in Sec. \ref{sect3subB} are satisfied, $n$-$\bar{n}$ oscillations are probably beyond the detectable regions of the present experiments.

The above results are obtained based on various assumptions and requirements, mainly from considerations regarding the type-\RomanNumeralCaps{2} seesaw mechanism and the naturalness of the vertex coupling constants. If we, however, loosen some of these requirements and assume that neutrino acquires Majorana masses directly from spontaneous breaking of the $(B-L)$ symmetry, then $n$-$\bar{n}$ oscillations may be accessible for the present experiments but the price we pay for such an assumption is an appropriate treatment of the problems arising from it, such as the existence of the massless Majoron particle. For example, if we ignore Eq. (\ref{lq1}) and choose $\lambda \simeq 10^{-3}$, $g \simeq 10^{-3}$, $f \simeq 10^{-13}$, and $M_{\Delta} \simeq 2.35$ TeV, then we get the $n$-$\bar{n}$ oscillation time $\tau_{n\bar{n}} \gtrsim 6.3 \times 10^{8}$ s, which is much stronger than the present limit of the direct search in the ILL experiment \cite{baldo1994new}, but may still lead to detectable effects in the present experiments. Similarly, if we ignore Eq. (\ref{lq1}) and choose $\lambda \simeq 10^{-2}$, $g \simeq 10^{-2}$, $f \simeq 10^{-13}$, and $M_{\Delta} \simeq 8.76$ TeV, then we get the $n$-$\bar{n}$ oscillation time $\tau_{n\bar{n}} \gtrsim 1.7 \times 10^{8}$ s, which is more accessible to direct searches. Moreover, under this assumption, it is required that the breaking of the $(B-L)$ symmetry occurs spontaneously roughly at the energy scale of $\sim 1$ TeV, which is lower than the ones ($\gtrsim 10$ TeV) proposed in previous studies, without invoking the type-\RomanNumeralCaps{2} seesaw mechanism \cite{barbieri1981spontaneous,costa1982higgs}.

\section{Conclusion and Outlook}

To summarize, we have analyzed the connection and compatibility between $n$-$\bar{n}$ oscillations and $\nu n \rightarrow \bar{\nu} \bar{n}$ reactions described by the interactions based on the $SU(3)_c \times SU(2)_L \times U(1)$ symmetry model with additional Higgs triplets. We have considered two scenarios of interest, corresponding to whether the $(B-L)$ symmetry could be broken. In scenario A, since $(B-L)$ is conserved, all the reported $n$-$\bar{n}$ oscillation candidates are actually produced by the solar neutrinos in the scattering process $\nu n \rightarrow \bar{\nu} \bar{n}$. In scenario B where both $(B+L)$ and $(B-L)$ could be broken, only a small fraction of the reported $n$-$\bar{n}$ oscillation candidates are actually produced by the solar neutrinos in the scattering process $\nu n \rightarrow \bar{\nu} \bar{n}$. Comparing with the $n$-$\bar{n}$ oscillation process, the $\nu n \rightarrow \bar{\nu} \bar{n}$ reaction process is described by higher-dimensional operators, and thus the effects are strongly suppressed by appropriate powers of energy scale associated with new physics, causing the signal too small to be detectable. In both scenarios, we have shown that the present detectors listed in Tab. \ref{tabtwo} are unable to distinguish a $n$-$\bar{n}$ oscillation event from a $\nu n \rightarrow \bar{\nu} \bar{n}$ reaction event within the accessible energy range. Nevertheless, if any of the two processes is detected, there could be signal associated with new physics beyond the SM \cite{phillips2016neutron,cerdeno2018bl}.

In scenario A where $(B-L)$ is unbroken, we find that $\nu n\rightarrow \bar{\nu}\bar{n}$ reactions are dominated by the $pp$ and $^7$Be solar neutrinos. The possible future availability of detecting the low-energy solar neutrinos with energies from $200$ keV to $1.0$ MeV could offer an opportunity to carry out more detailed and sensitive studies of the $(B+L)$ violations. Moreover, although the constraint on the energy scale, which is model dependent, seems not very useful in this scenario, the constraint on the cross section of the $\nu n \rightarrow \bar{\nu} \bar{n}$ reaction process is highly non-trivial. For example, the derived bound on the cross section at the average neutrino energy from the Super-K data is around $6.0 \times 10^{-51}$ cm$^2$, which is much smaller than the ones given by the typical electroweak and some non-standard neutrino-nucleon interactions \cite{papoulias2015standard,papoulias2019recent,sharma2018status}. A reasonable interpretation of such results requires further phenomenological studies using an appropriate effective model. Comparing with the solar neutrinos, the contribution to $\nu n\rightarrow \bar{\nu}\bar{n}$ reactions from other neutrino sources, such as the cosmic neutrino background, the supernova neutrinos etc., are not significant.

In scenario B where both $(B+L)$ and $(B-L)$ could be broken, we find that $\nu n \rightarrow \bar{\nu} \bar{n}$ reactions can serve to provide an additional constraint on the masses of the Higgs triplet bosons. Moreover, the prediction power can be greatly improved by comparing it with $n$-$\bar{n}$ oscillations in a way similar to Refs. \cite{mohapatra1982hydrogen,mohapatra1983spontaneous,mohapatra1983higgs} due to the elimination of large degree of uncertainty. We are interested in the appealing scenario where the mass of the Higgs triplet bosons is in the several TeV scale ($1$ TeV $\lesssim M_{\Delta} \lesssim$ $10$ TeV), which is accessible to a direct detection at the LHC or future high-energy experiments \cite{mohapatra2008diquark,dev2017naturalness,nomura2018discriminating,nomura2018discriminating,du2019type,de2019doubly}. We have explored the interplay of various requirements on the parameter space mainly in the type-\RomanNumeralCaps{2} seesaw framework. It is expected that if all these requirements are satisfied, the parameter space would be severely constrained. Our parameter scan shows that, in order to satisfy all the requirements listed in Sec. \ref{sect3subB}, the vertex coupling constant ($\lambda \simeq g \simeq f$) is roughly restricted in the range from the order of $10^{-3}$ to the order of $10^{-2}$. With the help of the acceptable parameters, we have estimated the bounds on the masses of the Higgs triplet bosons and have discussed their accessibility for a direct detection at the LHC or future high-energy experiments. The derived bounds on the masses of the Higgs triplet bosons from various experiments are approximately in the range from $2.4$ TeV to $8.8$ TeV, corresponding to two different scenarios with the vertex coupling constant $10^{-3}$ and $10^{-2}$ respectively. The derived bounds from different experiments are very close to each other and only weakly depend on the reported number of candidates, due to the fractional power dependence of Eq. (\ref{lq1}). If all the requirements are satisfied, although the masses of the Higgs triplet bosons could be within the reach of a direct detection at the LHC or future high-energy experiments, the predicted $n$-$\bar{n}$ oscillation times would be completely beyond the detectable regions of the present experiments. If we, however, loosen some of these requirements and assume that neutrino acquires Majorana masses directly from spontaneous breaking of the $(B-L)$ symmetry, then $n$-$\bar{n}$ oscillations may be accessible in the present experiments but the price we pay is an appropriate treatment of the problem arising from such an assumption, for example, the existence of the massless Majoron particle.

\section*{Acknowledgement}
I would like to thank Prof. Rob Timmermans and Dr. Anastasia Borschevsky for providing help and support, and thank Femke Oosterhof and Ruud Peeters for many useful conversations.

\bibliography{NNBar}

\end{document}